\documentclass[prl,floatfix,twocolumn,showpacs,preprintnumbers,amsmath,amssymb,superscriptaddress]{revtex4}
\usepackage{graphicx,psfrag}
\usepackage{amsfonts}

\newcommand{\newc}{\newcommand}
\newc{\gsim}{\lower.7ex\hbox{$\;\stackrel{\textstyle>}{\sim}\;$}}
\newc{\lsim}{\lower.7ex\hbox{$\;\stackrel{\textstyle<}{\sim}\;$}}
\newc{\gev}{\,{\rm GeV}}
\newc{\mev}{\,{\rm MeV}}
\newc{\ev}{\,{\rm eV}}
\newc{\kev}{\,{\rm keV}}
\newc{\tev}{\,{\rm TeV}}

\newc{\mz}{M_Z}
\newc{\mpl}{M_*}
\newc{\mw}{m_{\rm weak}}
\newc{\nr}[1]{N^c_R{}_{#1}}
\usepackage{amsmath}
\def\beq{\begin{equation}}
\def\eeq{\end{equation}}
\def\bea{\begin{eqnarray}}
\def\eea{\end{eqnarray}}
\def\bitem{\begin{itemize}}
\def\eitem{\end{itemize}}
\newc{\ie}{{\it i.e.}}          \newc{\etal}{{\it et al.}}
\newc{\eg}{{\it e.g.}}          \newc{\etc}{{\it etc.}}
\newc{\cf}{{\it c.f.}}
\def\inv{^{\raise.15ex\hbox{${\scriptscriptstyle -}$}\kern-.05em 1}}
\def\lbar{{\lower.35ex\hbox{$\mathchar'26$}\mkern-10mu\lambda}} 

\let\<=\langle
\let\>=\rangle

\let\+=\uparrow

\let\De=\Delta

\let\Si=\Sigma
\let\th=\theta

\begin{document}
\thispagestyle{empty}

\begin{center}

\title{Kerr Black Holes as Particle Accelerators to Arbitrarily High Energy}
\date{21st August 2009}
\author{M\'aximo Ba\~nados}
\email{maxbanados@fis.puc.cl}
\affiliation{Facultad de F\'{\i}sica, Pontificia Universidad Cat\'{o}lica de Chile, Av. Vicu\~{n}a Mackenna 4860, Santiago, Chile }
\affiliation{Physics Department, University of Oxford, Oxford, OX1 3RH, UK}
\author{Joseph Silk}
\email{j.silk1@physics.ox.ac.uk}
\affiliation{Physics Department, University of Oxford, Oxford, OX1 3RH, UK}
\author{Stephen M. West}
\email{stephen.west@rhul.ac.uk}
\affiliation{Royal Holloway, University of London, Egham, TW20 0EX, UK}
\affiliation{Rutherford Appleton Laboratory, Chilton, Didcot, OX11 0QX, UK}

\begin{abstract}
We show that intermediate mass  black holes conjectured to be the early precursors of supermassive black holes and  surrounded by   relic cold dark matter density spikes can act as particle accelerators with collisions, in principle, at arbitrarily high centre of mass energies  in the case of Kerr black holes. While the ejecta from such interactions will be highly redshifted, we may anticipate the possibility of a unique probe of Planck-scale physics.  \end{abstract}
\maketitle

\end{center}


In this Letter, we show that rotating black holes may act as particle accelerators. The context we have in mind is that of intermediate mass  black holes conjectured to be the early precursors of supermassive black holes and  surrounded by   relic cold dark matter density spikes. Collisions between particles, e.g. massive dark matter particles, may reach arbitrarily high center-of-mass  energies. Naively speaking this may not come as a surprise because particles are infinitely blue shifted near the horizon (for an observer sitting there). However, the center of mass between two particles is a free fall frame and the energy in this system is bounded. For Schwarzschild black holes, the maximum energy is
$E^{max}_{cm} = 2\sqrt{5}\, m_0,$
where $m_0$ is the mass of the two colliding particles  \cite{Baushev}. Here it is assumed that the particles at infinity are at rest, and the collision energy comes solely from gravitational acceleration.  Note that this limit does not depend on the mass of the black hole. 

Since it is expected that most astrophysical black holes would have large angular momenta, in fact near extremality \cite{Thorne}, it is natural to ask what is the maximum center of mass energy for Kerr black holes.  After computing $E_{cm}$ for Kerr we show that the maximum energy grows with $a={J \over M}$.  Furthermore, as the black hole becomes extremal, $E^{max}_{cm}$ grows without limit providing an accelerator that in principle  allows collisions at arbitrarily high energies. We concentrate here on our main new result, namely computation of  the limiting energy for Kerr black holes. The energy distribution of particles colliding at this maximum energy will be discussed in a future publication.

\begin{figure}
\includegraphics[width=9cm]{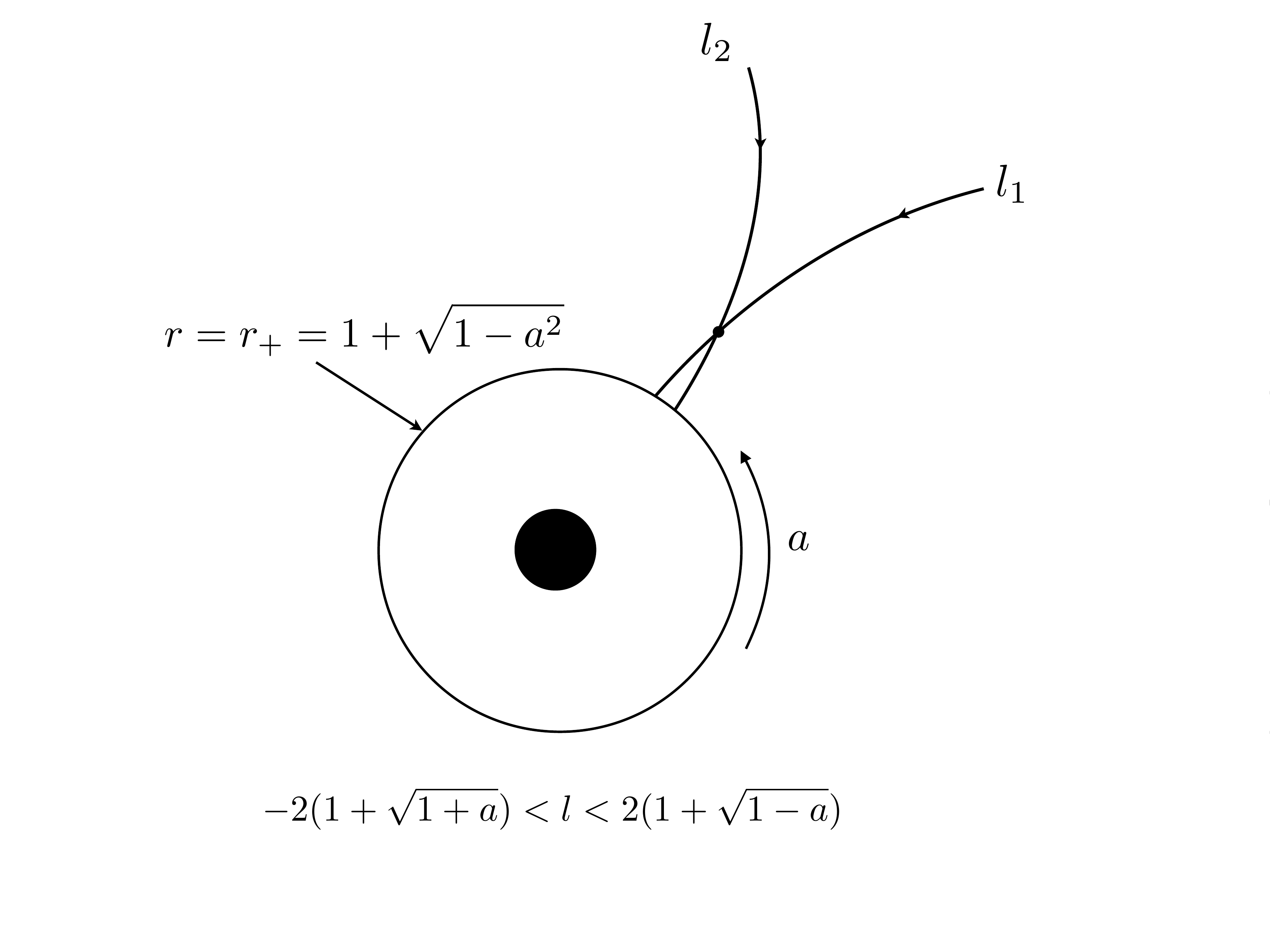}
\vspace{-10mm}
\caption{\label{fig0}Schematic picture of two particles falling into a black hole with angular momentum $a$ (per unit black hole mass) and colliding near the horizon. The allowed range of $l$ for geodesics falling in to the black hole is also given.}
\vspace{-10mm}
\end{figure}

The general situation we consider in this paper is depicted in Fig. \ref{fig0}. Two particles are falling into the black hole and collide near the horizon. The range of $l$, the angular momentum per unit rest mass, for geodesics falling in is $-2(1+\sqrt{1+a}) \leq l \leq 2(1+\sqrt{1-a})  $  (see \cite{BPT} for a detailed treatment of geodesics in Kerr backgrounds). We shall not deal with the global properties of the geodesics. The figure represents the local properties  before the collision.
We  start by describing the situation for Schwarzschild black holes, as described in \cite{Baushev}. Our treatment will be general enough such that the extension to Kerr will be almost direct.
Consider two particles approaching the black hole with different angular momenta $l_1$ and $l_2$ and colliding at some radius $r$ (see Fig. \ref{fig0}). Later, we consider the collision point $r$ to approach the horizon, $r=r_+$. The particles will be assumed to be at rest at infinity. 
Non-rotating black holes are described by the Schwarzschild metric (where we have set the mass of the black hole to 1),
\begin{equation}
ds^2 = -\left(1-{2 \over r}\right)dt^2 + \left(1-{2 \over r}\right)^{-1}dr^2 + r^2(d\theta^2+\sin^2\theta d\varphi^2)
\end{equation}
and the solution to the geodesic equation is (see \cite{BPT} for a detailed analysis of geodesics on black hole spacetimes)
\begin{eqnarray}
  {dt \over d\tau} &=& \left( 1- {2 \over r} \right)^{-1},   \nonumber\\
  {dr \over d\tau} &=& -{1 \over r^2}\sqrt{r(2r^2 + 2l^2 -  rl^2 ) }, \nonumber\\
  {d\varphi \over d\tau} &=& {l \over r^2}. \label{geos}
\end{eqnarray}
We assume throughout the paper that the motion of particles occur in the Equatorial plane.
\begin{figure*}
\centerline{\includegraphics[width=15cm,height=6cm]{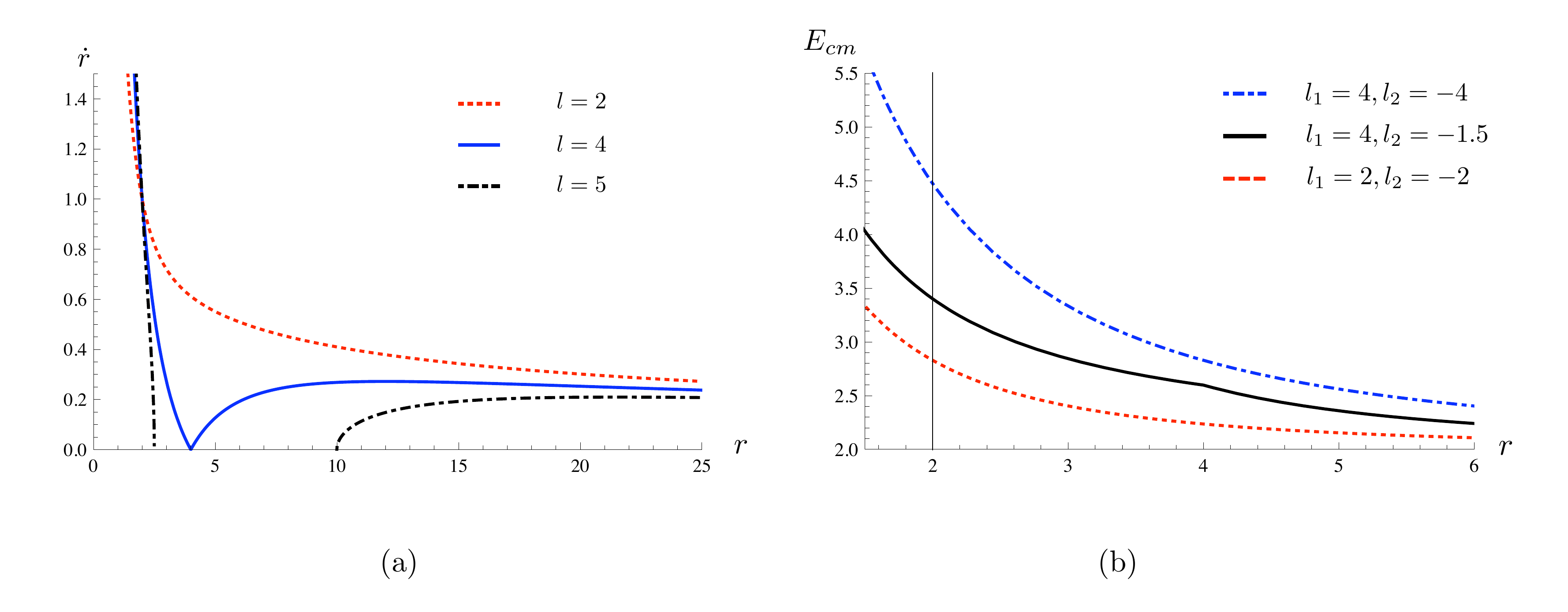}}
\caption{For a Schwarzschild black hole (a) shows the variation of $\dot{r}$ with radius for three different values of angular momentum $l=2, 4$ and $5$. For $l=5$ $\dot r$ reaches zero, indicating a turning point, before reaching the horizon. These geodesics are not interesting for us. $l=4$ is a critical case where geodesics start falling in. For smaller values of $|l|$ all geodesics fall in. Panel (b) shows the variation of $E_{cm}$ with radius for three combinations of $l_1$ and $l_2$. Note the kink corresponding to the critical geodesic with $l=4$. The vertical line at $r=2$ is the horizon and we see that $E_{cm}$ is finite in all cases. }
\label{fig1}
\end{figure*}
The maximum center of mass energy arises when the particles approach the black hole from opposite directions and maximum (opposite) angular momentum. There are two competing effects. If the angular momentum is too large (see Fig. (\ref{fig1}a)), geodesics never reach $r=r_+$  and the near horizon physics is not probed. On the other hand, if the angular momentum is too small,  the particles fall radially with a small tangential  velocity and the center of mass energy does not grow either. Consequently, there is a critical value for the angular momentum such that  particles reach the horizon with maximum tangential velocity. For a Schwarzschild black hole, the critical values are $l=\pm 4$.

Our aim is to compute the energy in the center of mass frame for this collision. Since the background is curved, we need to define the center of mass frame properly. It turns out that there is a simple formula for $E_{cm}$ valid both in flat and curved spacetimes,
\begin{equation}\label{Ecm}
E_{cm} = m_0\sqrt{2} \sqrt{1 - g_{\mu\nu} u^{\mu}_{_{(1)}} u^\nu_{_{(2)}}  },
\end{equation}
where $u^\mu_{_{(1)}}$ and $u^\nu_{_{(2)}}$ are the 4-velocities of the particles, properly normalized by $g_{\mu\nu}u^\mu u^\nu=-1$ (we use the mostly plus signature). This formula is of course well-known in special relativity, and the principle of equivalence should be enough to ensure its validity on a curved background. 

The derivation of (\ref{Ecm}) on curved backgrounds can be done in an efficient way by using the notion of orthonormal frames introduced in \cite{BPT}. At any point one can set a `lab' frame described by vectors $\hat {e}_a$ satisfying $\hat e_a \cdot \hat e_b = \eta_{ab}$. The basis vectors $\hat{e}_a$ and the coordinate basis $\vec{e}_\mu$ are related by an invertible matrix $e^{a}_{\ \mu}$ such that $\vec{e}_\mu = e^{a}_{\ \mu} \hat{e}_a$.  If the metric $g_{\mu\nu}$ is known, then the matrix $e^{a}_{\ \mu}$ is uniquely defined, up to a Lorentz transformation. Given the worldline history $x^\mu(\tau)$ and the 4-velocity $u^\mu = {dx^\mu \over d\tau},$ we can construct the 3-velocities observed in the lab frame \cite{BPT},
\begin{equation}\label{}
v^{(i)} = { e^{(i)}_\mu u^\mu \over e^{(0)}_{\mu} u^\mu }.
\end{equation}
In this frame special relativity holds, thus, the components of the momentum of  a particle of mass $m_0$ are given by
\begin{equation}
p^a = m_0 \gamma(v) (1,v^{(i)}),
\end{equation}
where $\gamma(v)= (1-v^2)^{-1/2}$. In this frame, the center of mass energy formula (\ref{Ecm}) holds with $g_{\mu\nu}$ replaced by $\eta_{ab}$. One can now directly go from the flat space version of (\ref{Ecm}) into a curved background simply by using the transformations given above, in particular $g_{\mu\nu}=e^{a}_{\ \mu}e^{b}_{\ \nu}\eta_{ab}$.

We now apply (\ref{Ecm}) to the problem of two particles colliding in the Schwarzschild background. By direct replacement of  (\ref{geos}) into (\ref{Ecm}) one finds,
\begin{widetext}\
\beq
\label{ESch}
{1 \over 2m_0^2} \Big( E_{cm}^{^{Schw}} \Big)^2 = { 2 r^2(r -1) - l_1 l_2(r-2) - \sqrt{2r^2 -l_1^2(r-2)}\sqrt{2r^2 - l_2^2(r-2)}  \over r^2(r-2)}.
\eeq
\end{widetext}

The horizon is located at $r=2$ and we naively observe a pole ${1/ (r-2)}$ which seems characteristic of an infinity blueshift.  However, this is not the case as the numerator also vanishes at $r=2$. In fact the limit is finite and equals
\begin{equation}\label{lim2}
E_{cm}^{^{Schw}}(r\rightarrow 2) = {m_0 \over 2} \sqrt{(l_2-l_1)^2 + 16}.
\end{equation}
The maximum center of mass energy occurs for $l_2$ and $l_1$ opposite with their maximum allowed values (for geodesics falling in to the black hole) $l_1=4$ and $l_2=-4$. Using these values in (\ref{lim2}) one obtains $2\sqrt{5}m_0$, as stated in the first paragraph \cite{Baushev}. Note also that if $l_2=l_1$, $E_{cm}=2m_0$ as it should be. A finite limit in the center of mass energy follows from the fact that all particles approach the horizon with the same incident angle (that is they approach perpendicularly) and thus their relative velocities go to zero.
\begin{figure*}
\centerline{\includegraphics[width=15cm,height=6cm]{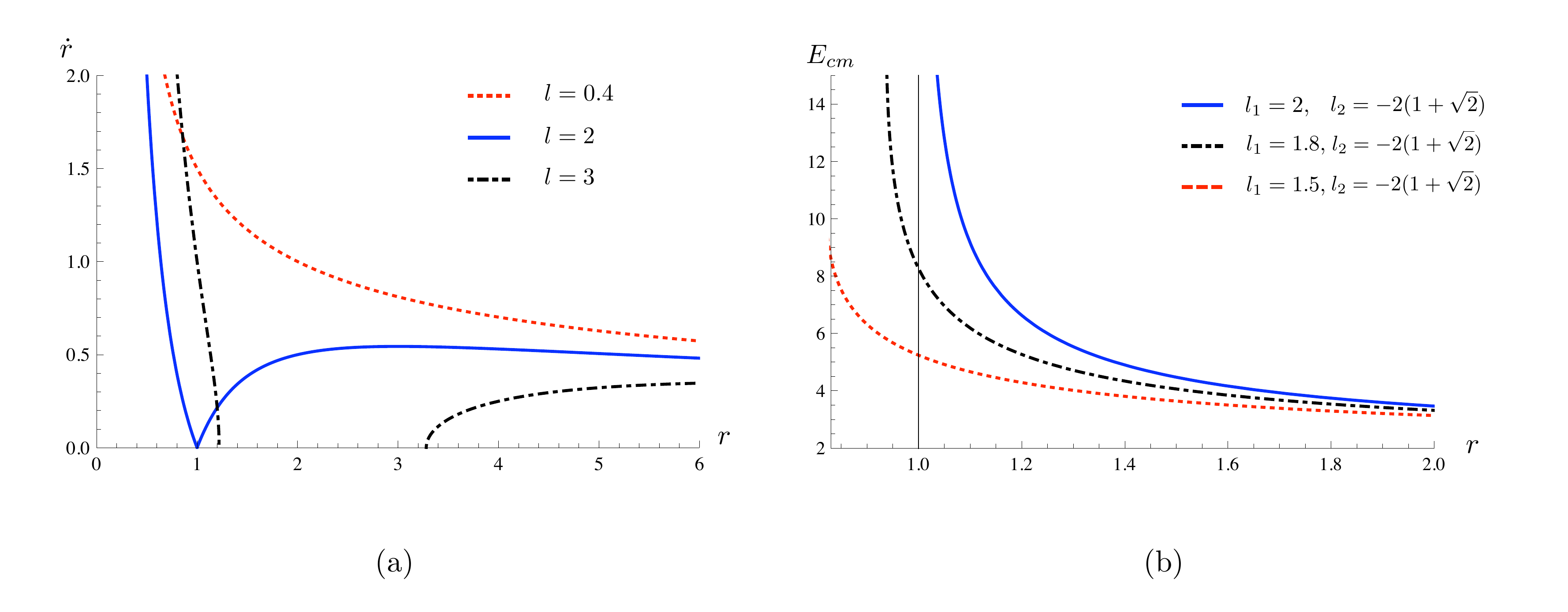}}
\caption{\label{fig3}  For a Kerr Black hole with $a=1$ (a) shows the variation of $\dot{r}$ with radius for three different values of angular momentum $l=0.4, 2$ and $3$.  (b) shows the variation of $E_{cm}$ with radius for three combinations of $l_1$ and $l_2$. For $l_1=2$ we see that $E_{cm}$ blows up at the horizon.}
\end{figure*}
We shall now apply (\ref{Ecm}) for particles moving on a Kerr black hole. We again restrict
the discussion to equatorial geodesics.  The extension to more general geodesics has no conceptual problems but the formulae involved get rather cumbersome. Our main goal here  is to show that on a Kerr black hole particles may collide with arbitrarily large center of mass energy, and this can already be seen on equatorial geodesics. Again, we assume that the particles are at rest at infinity, and that all collision energies are provided by gravitational acceleration.

The Kerr black hole is described by the metric (where again we have set the mass of the black hole to 1)
\begin{multline}
ds^2=-\left(1-\frac{2r}{\Si}\right)dt^2-\left(\frac{4ar\sin^2\th}{\Si}\right)dtd\varphi\\ + \left(\frac{\Si}{\De}\right)dr^2+\Si d\th^2 +\left(r^2+a^2+\frac{2a^2r\sin^2\th}{\Si}\right)\sin^2\th d\varphi^2
\end{multline}
where $a$ is its angular momentum per unit mass ($0\leq a \leq 1$) and the functions $\De$ and $\Si$ have the forms
\bea
\De&\equiv& r^2+a^2-2r\\
\Si &\equiv & r^2 + a^2 \cos^2\theta
\eea
Let $x^{\mu}(\tau)$ be the history of a particle of mass $m_0$. The geodesic equation can be fully integrated (Carter \cite{Carter:1968}). The equations of motion governing the orbital trajectory of a particle on the equatorial plane, $\theta=\pi/2$, read (see \cite{BPT})
 \bea
\frac{d r}{d \tau}&=& \pm \frac{1}{r^2}\sqrt{T^2-\De(m_0^2r^2+(l-aE)^2)},\\
\frac{d \varphi}{d \tau}&=& -\frac{1}{r^2}\left[\left(aE-l\right)+aT/\De\right],\\
\frac{d t}{d \tau}&=& -\frac{1}{r^2}\left[ a(aE - l)+(r^2+a^2)T/\De\right],
\eea
 where $ T \equiv E(r^2+a^2)-la$. Here $E$ is the total energy of the particle and $l=p_{\varphi}$ is the component of angular momentum parallel to the symmetry axis per unit mass.

On a Kerr background, particles approaching from one or the other side have different properties.  Consider two particles coming from infinity with $E_1/m_0=E_2/m_0=1$ and approaching the black hole with different angular momenta $l_1$ and $l_2$.    By direct application of (\ref{Ecm}) we find the generalization of (\ref{ESch}) for rotating backgrounds,
\begin{widetext}
\begin{multline}
\Big(E^{^{Kerr}}_{cm}\Big)^2=\frac{2\, m_0^2}{r (r^2 - 2r + a^2)} \Big(  2 a^2 (1 + r)-2 a (l_2 + l_1) - l_2 l_1 (-2 + r) + 2 (-1 + r) r^2   \\  -\sqrt{2 (a - l_2)^2 - l_2^2 r + 2 r^2} \sqrt{2 (a - l_1)^2 - l_1^2 r + 2 r^2}\, \Big).\label{EcmK}
\end{multline}
\end{widetext}
The horizon is located at the larger root of $r^2-2r + a^2=0$, that is $r_{+} = 1 + \sqrt{1-a^2}$. As for the Schwarzschild case, it appears that $E^{^{Kerr}}_{cm}$ diverges at $r=r_+$ but again this is not true because, although not totally evident, the numerator vanishes at that point as well. The limiting value of $E^{^{Kerr}}_{cm}$ at the horizon for generic values of $a$ is not very illuminating. However, in the case that $a=1$ the form of the centre of mass energy at $r_+$ reads
\begin{equation}\label{limK}
E^{^{Kerr}}_{cm}(r\rightarrow r_+) = \sqrt{2}m_0 \sqrt{ {l_2 -2 \over l_1-2} + {l_1 -2 \over l_2-2}  }, \ \ \ \ \  (a=1)
\end{equation}
which is indeed finite for generic values of $l_1$ and $l_2$. Observe that for $a=1$ the denominator in (\ref{EcmK}) has a double zero and it can also be shown that the numerator has a double zero also. This is true except when $l=2$.  A new phenomena appears if one of the particles participating in the collision has the critical angular momentum $l=2$.  We see that in this case the limit (\ref{limK}) ceases to exist, and in fact the center of mass energy blows up at the horizon.

We plot in Fig. \ref{fig3}(b) $E_{cm}^{^{Kerr}}(r)$ for various values of $l_1$ and $l_2$. Observe that the critical value $l=2$ corresponds to particles coming from infinity with the maximum allowed angular momentum for a geodesic falling into the black hole.  This is shown in Fig. \ref{fig3}(a). As we mentioned before, a limit in the center of mass energy follows from the fact that all particles approach the horizon with the same incident angle (perpendicularly) and thus their relative velocities go to zero. This property fails for the critical geodesic with $l=2$ (at $a=1$). In this case, the particle hits the horizon non-perpendicularly and consequently has a non zero tangential velocity allowing for large center of mass energy collisions.

In this Letter, we have analysed the possibility that Kerr black holes can act as natural particle accelerators. It has previously been argued that the formation of central or isolated black holes within cold dark matter subhalos or halos by baryonic accretion, either by gaseous dissipation of an accretion disk or by stellar tidal disruption, is accompanied by generation of a central dark matter spike \cite{GS}. The maximum extent of the spike is determined by the gravitational radius, the density profile slope by the initial dark matter core profile, and the innermost extent where the spike flattens by the annihilation cross-section. Such spikes yield detectable annihilation signals in gamma rays, high energy neutrinos or antiparticles.

In these earlier works the discussions considered annihilations at rest. In this letter we note that some rare collisions may occur at arbitrarily high energies for the case of Kerr black holes. In effect we have a Planck scale particle accelerator. Of course the ejecta from such interactions will be highly redshifted. However, the possibility of a unique probe of high-scale physics via signals buried in the emerging debris is an intriguing prospect. No terrestrial accelerator could conceivably approach the energies that would be achievable in rare plunging orbits around a Kerr black hole. In a future paper, we will explore the flux and energy distribution of the collision debris for particular particle models. 

MB is a J.S. Guggenheim Memorial Foundation fellow and is partially supported by Fondecyt (Chile) Grant \#1060648 and Alma Grant \# 31080001.  SMW thanks the Oxford physics department for hospitality and the Higher Education Funding Council for England and the Science and Technology Facilities Council for financial support under the SEPNet Initiative.


\begin{thebibliography}{1}

\bibitem{Baushev}
A.~N. Baushev.
\newblock {Dark matter annihilation in the gravitational field of a black
  hole}, arXiv:0805.0124 [astro-ph].
\bibitem{Thorne}
Kip~S. Thorne.
\newblock {Disk accretion onto a black hole. 2. Evolution of the hole}.
\newblock {\em Astrophys. J.}, 191:507, 1974.

\bibitem{BPT}
James~M. Bardeen, William~H. Press, and Saul~A Teukolsky.
\newblock Rotating black holes: Locally nonrotating frames, energy extraction, and scalar synchrotron radiation.
\newblock {\em Astrophys. J.}, 178:347, 1972.

\bibitem{Carter:1968}
Brandon Carter.
\newblock Global structure of the kerr family of gravitational fields.
\newblock {\em Phys. Rev.}, 174:1559--1571, 1968.

\bibitem{GS}
Paolo Gondolo and Joseph Silk.
\newblock {Dark matter annihilation at the galactic center}.
\newblock {\em Phys. Rev. Lett.}, 83:1719--1722, 1999.

\end{thebibliography}
\end{document}